# PlanetSense: A Real-time Streaming and Spatio-temporal Analytics Platform for Gathering Geo-spatial Intelligence from Open Source Data "(Vision Paper)"

Gautam S. Thakur, Budhendra L. Bhaduri, Jesse O. Piburn,
Kelly M. Sims, Robert N. Stewart, Marie L. Urban
The Geographic Information Science and Technology (GIST) Group
Oak Ridge National Laboratory, 1 Bethel Valley Road, Oak Ridge, TN 37831 USA
{thakurg, bhaduribl, piburnjo, simskm, stewartrn, urbanml}@ornl.gov

## ABSTRACT
Geospatial intelligence has traditionally relied on the use of archived and unvarying data for planning and exploration purposes. In consequence, the tools and methods that are architected to provide insight and generate projections only rely on such datasets. Albeit, if this approach has proven effective in several cases, such as land use identification and route mapping, it has severely restricted the ability of researchers to inculcate current information in their work. This approach is inadequate in scenarios requiring real-time information to act and to adjust in ever changing dynamic environments, such as evacuation and rescue missions. In this work, we propose *PlanetSense*, a platform for geospatial intelligence that is built to harness the existing power of archived data and add to that, the dynamics of real-time streams, seamlessly integrated with sophisticated data mining algorithms and analytics tools for generating operational intelligence on the fly. The platform has four main components – i) GeoData Cloud – a data architecture for storing and managing disparate datasets; ii) Mechanism to harvest real-time streaming data; iii) Data analytics framework; iv) Presentation and visualization through web interface and RESTful services. Using two case studies, we underpin the necessity of our platform in modeling ambient population and building occupancy at scale.

## Categories and Subject Descriptors
C.3 [**Special Purpose and Application Based Systems**]: Real time, Distributed Systems. H.2.8 [**Database Applications**]: Data mining, Image databases, spatial databases and GIS.

## 1. INTRODUCTION
Geospatial intelligence has a long history of exploring and analyzing geospatial information to describe geographically referenced activities and features on the planet. The exploration is the result of extracting meaningful information and discovering hidden facts from imagery, sensors, and other geospatial data. In several cases, discoveries are made using all three capabilities or a combination of them. In addition to the spatial reference, geospatial intelligence incorporates temporal information and provides insight through visualization. Various values added and baseline statistics on infrastructure, weather, Internet of Things (IoT) data, medical, and logistics add context to geospatial analysis. The information generated through geospatial intelligence has application in several important areas including aeronautics, land use, urban systems, national security, emergency management, and topographical mapping.

Current approaches to gathering intelligence largely depend on the data collected some time in the past (e.g. surveys, census). The focus is on the use of archived facts for generating deep analysis and predictions. The tools and methods are also designed to work on such data sources. While acting based on past data is effective, it entirely misses the opportunity to harvest real-time data and instantly act on it. Several open source datasets, satellites, social media, and sensors are capable of monitoring real-time geospatial event data. Thus, operating on such data and instantly communicating information can aid evacuation management, estimate ambient population, adaptive traffic management, or contain the spread of wildfires. It also enables accurate estimation and management of required resources as events happen. While one cannot deny the importance of archived data with the pursuit of gathering intelligence, the addition of real-time data will allow for a level of awareness and responsiveness to real-time developing events that archived data alone cannot provide.

In this work, we vision to design a scalable platform that extends existing geospatial data with real-time streaming data, collected from numerous open sources, social media, and passive and participatory sensors (IoTs, traffic cameras, detectors). We contend that such a platform must encompass a range of technologies and tools to generate responsive intelligence in real-time. The storage and processing requirements span across multiple distributed machines, seamlessly integrating streaming data with unvarying datasets. We witness the need for a big data architecture surpassing the capabilities for traditional approaches, relying on hardware and software components, and supporting larger and faster processing mechanisms. In addition, the platform exploits machine learning and data mining methods to act on the data and extract information on the fly. In order to achieve these design goals, we propose *PlanetSense*, a real-time streaming and spatio-temporal analytics platform for discovering geospatial intelligence. This distributed and scalable platform aims to follow big data principles, while the actual deployment is tailored for geospatial work. The platform has ability to combine large volumes of historic and real-time data at the same time. A novel feature is an analytics layer that input data for discovering intelligence in real-time. A set of visualization and easy to understand interfaces provide an interactive environment for end users. Section 2 discusses related work and platform's architecture is explained in Section 3. Two case studies based on a working prototype are discussed in Section 4 and conclusion in Section 5.

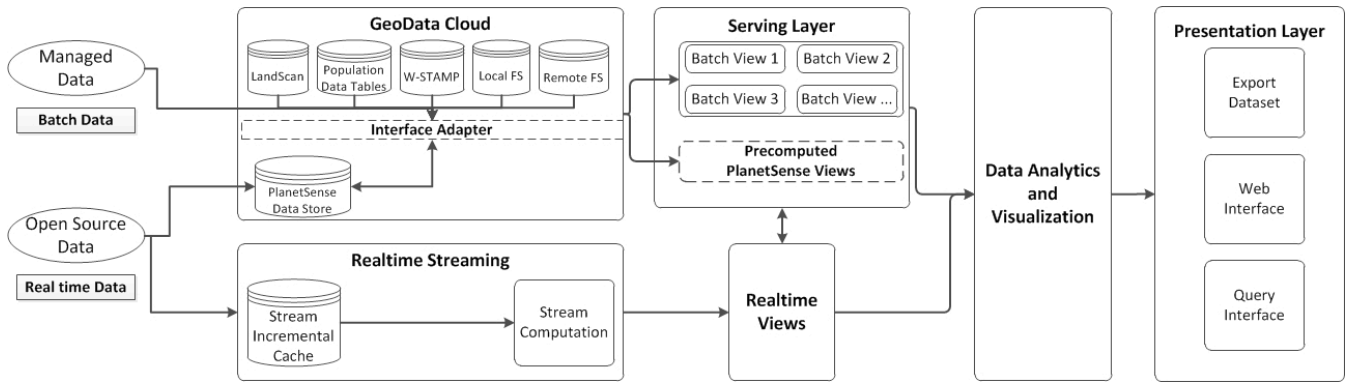

**Figure 1 PlanetSense Architecture**

## 2. RELATED WORK

Recently, intelligence agencies have expressed interest in answering key intelligence questions using only unclassified data such as open source and social media data[1]. This type of data is mainly collected in two different forms. The first one is based on crowdsourcing and Volunteered Geographical Information (VGI), with citizens as sensors collecting and contributing geo-spatial data to the community[2], [3]. The second approach is Ambient Geographical Information (AGI), where the citizens serve as data points and provide a better understanding on human landscape[4]. AGI data is mainly contributed through the use of online social networks, such as Twitter and Facebook, where the users of these tools share responses, concerns, information etc. that can provide an additional layer to the traditional GIS data sources. Among these, social media data (e.g. Twitter) or IoTs also generate real-time streaming information. To realize the use of such data, one important criterion is the robust and responsive architecture that will support the collection, management, processing and retrieval of very large data frames in a matter of a few seconds. Several architectures were proposed in the past decade supporting various forms of streaming data processing[5]–[8]. However, their focus was entirely on the streaming data collection and made no efforts to incorporate archived data. Our architecture supports a seamless integration of archived as well as streaming data. Some social media systems such as GeoSocial Gauge and TaghReed minimally support data analytics and exploratory statistical analysis[9], [10]. One of the important aspects of our platform is not only the collection and management of disparate data sources, but also the ability to perform complex mining and machine learning tasks on the data before it is presented to the end user.

## 3. PLANETSENSE PLATFORM

The PlanetSense architecture is the integration of several technologies and tools that builds the big data collection and processing platform. The focus is towards integrating disparate data sources, real-time analytics, and rich visualization support.

### 3.1 Motivation

Over the years, the GIST group at ORNL has produced several data products (discussed later in section 3.2.2), focusing on ambient population dynamics, world spatio-temporal analytics, and population density tables, respectively. We want to bring streaming data capability to these immutable data products to improve our understanding of geospatial information and capture the dynamic nature of human populations. Another motivation comes from the need to provide a mechanism to perform real-time analytics on the data to generate operational intelligence on the fly. Initially, we started using a standard configuration of MySQL database server and single threaded webserver, but soon encountered difficulties in harvesting enormous quantities of streaming data from social and open source media. So we decided to harness the power of big data architectures that provide support for scaling and extensibility. We considered these issues and concluded that our best option is to vision a new streaming architecture from the ground up that combines existing data stores with streaming sources and provides the ability to perform real-time data analytics.

### 3.2 Architecture

In this section, we briefly describe the PlanetSense architecture, geospatial data store - "GeoData Cloud", GIS workflow, and various components of PlanetSense as shown in Figure 1. Our knowledge of Lambda architecture proved vital in designing this new system[8]. We have addressed two major challenges dealing with legacy geospatial data (mostly relational) and processing requirements for real-time operational intelligence.

#### 3.2.1 Data Input Model

There are two types of data source entering into the system. The first are unvarying data loads that consist of surveys (e.g. census data) only periodically updated. The second type is the real-time streaming data entering from social media, imagery, open data sources (Wikipedia, traffic signals, etc.), and Internet of Things (IoT) mostly based on geographical preference. The streaming data entering in the system is then dispatched both to the GeoData Cloud and real-time streaming layer for further processing.

#### 3.2.2 GeoData Cloud

The GeoData cloud hosts disparate and archived spatio-temporal datasets and an incremental copy of real-time streamed data for future purposes in a separate datastore. Some spatio-temporal datasets such as LandScan provide ambient population statistics[11], W-STAMP provides spatiotemporal information gleaned from public sources like the CIA World Factbook[12],

Population Density Tables (PDT) project developing population density estimates for specific human activities under normal patterns of life based largely on information available in open source[13]. These dataset are independently owned and managed, PlanetSense access them via *read-only* interface adapters.

### 3.2.3 Real-time Streaming Layer
The purpose of real-time streaming layer is to compensate for GeoData Cloud's access latency and lagged processing time. It has a small data footprint and works by first caching the streaming data entering from outside sources and then processing minimally, generating real-time information. Currently, our prototype allows harvesting streaming data from several major social networks and open source sites such Wikipedia, weather, and traffic cameras.

### 3.2.4 Serving Layer
The serving layer is responsible for two activities – (i) Creating batch views, which are based on the immutable geographical spatio-temporal data (E.g. specific LandScan views), (ii) Pre-computed PlanetSense views, which are based on earlier archived stream data. The pre-computed geospatial views are generated for augmenting the result of streaming results. For example, a pre-computed view consists of land use (commercial, residential, and mix) to identify the list of businesses in a geographical bounding box, and use ambient population in that area through streaming analysis based on social media check-in counts. It is natural for the data that resides in the cloud to grow over time; accordingly the batch views are recomputed continuously from scratch.

### 3.2.5 Real-time Views
The real-time views are computed for the data that are currently streaming into the system and not represented in the batch view system yet. The purpose of the real-time views is to use incremental updates as they arrive through the streaming sources.

### 3.2.6 Data Analytics and Visualization
The PlanetSense architecture comes with an array of sophisticated machine learning and data mining algorithms to gather operational intelligence. These algorithms consume the processed views from the batch and streaming layer into the merged layer. The merged layer also stores trained machine learning models such as topic models that extract scenarios based on data collected through searches. The visualization library complements analytics and serves as a high-performance 2D/3D visualization platform.

### 3.2.7 Presentation Layer
The presentation layer provides users multiple interfaces to interact and gather geospatial intelligence on the fly. There are currently three different interfaces provided based on the kind of operation required for the information generation.

#### 3.2.7.1 Export Dataset
Users wanting to store and use the data can download the processed data containing the relevant information.

#### 3.2.7.2 Web Interface
The web interface is designed to provide inference based learning extracted from the analytics layer. Also, the results of the algorithms are modeled and plotted through the interface layer.

#### 3.2.7.3 Query Interface
This is an interactive interface that provides ad hoc execution of user queries. This interface provides succinct information in form of text data output to the end user.

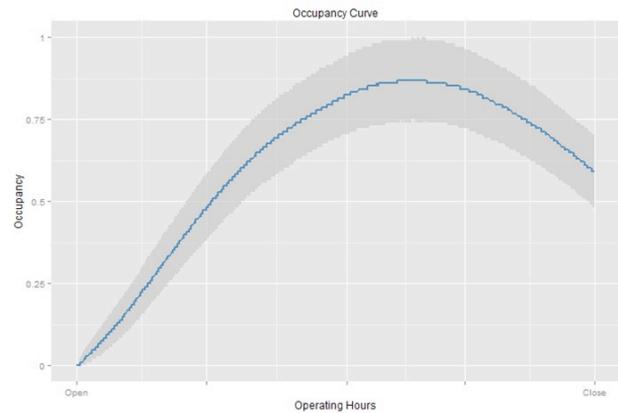

**Figure 2 High Museum Unit Signatures**

## 4. CASE STUDY
In this section, we showcase the usability of our prototype platform to examine occupancy in museums and dynamics of special event population using real-time streamed social data.

### 4.1 Occupancy
With the increase in the volume of available data, small area population estimates are advancing towards ever finer spatial resolutions. The Population Density Tables (PDT) project takes this down to the scale of estimating the occupancy at the granularity of actual building and facility spaces. Average day and night estimates are derived from a Bayesian machine-learning model that is updated by newly collected open source data and subject matter expertise. While this alone enables fine spatial resolution, to attempt an approach at a finer temporal resolution, ancillary data with a higher velocity is needed.

The use of social media data is a potential source for providing a finer temporal resolution to building occupancy estimates. A first attempt at this methodology was to use Facebook Check-Ins at the High Museum in Atlanta, GA to estimate a unit occupancy curve for the museum's operating hours. Through the use of a unit occupancy model that informs prior PDT estimates with social media data a probabilistic occupancy signature was derived. Figure 2 shows the unit occupancy estimates along with 95% confidence intervals. For a detailed study, read[14].

### 4.2 Mapping of a Special Event Population
In this case study, we use social media data for modeling the dynamic nature of human populations. A natural application of the geo-located subset of social media data is the modeling of episodic populations associated with special events with high attendance and a significant presence on social media; in particular, this study focuses on game day college football fans at The University of Tennessee - Knoxville. Using tweets and Facebook check-ins, this research integrates this new form of data in a high-resolution dasymetric population distribution model. The area within a 1.5-mile radius around The University of Tennessee football stadium was chosen for this study, and the population associated with football game days was modeled. Geo-located tweets and check-ins were collected for the 24-hour period surrounding the scheduled kickoff for each home game in 2013. A cumulative count of check-ins was captured every 30 minutes for 95 establishments associated with game day activities (e.g., restaurants and tailgating locations).  Two scenarios were modeled: 1) a "non-game-hours" scenario and 2) a "game-hours" scenario. Each model outputs a population estimate for each cell in a raster grid with 3 arc-second resolutions (~90 m). The 2012 version of the LandScan USA[11] gridded nighttime dataset was used as a baseline population distribution to which the new-modeled distributions could be added to create the final output grids. A count of tweets and a count of check-ins were computed for each raster cell for each of the two scenarios, resulting in four raw count rasters. Kernel density estimation with a radius of two grid cells was performed on each of the four raw count rasters to estimate tweet densities and check-in densities across the grid. High populations can be seen in and near the stadium in both, but with much greater concentration in Figure 3(b). Figure 3(a) shows greater concentrations in areas on and near campus that are popular for tailgaters as well as along Cumberland Avenue and in the downtown area, where restaurants, bars, and shops are concentrated. For detailed analysis, read[15].

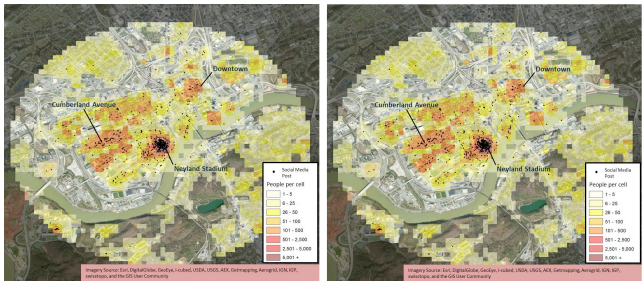

**Figure 3 Example game day population distribution. a) During non-game-hours. b) During game hours**

## 5. CONCLUSIONS AND FUTURE WORK

The glut of real-time volunteered data sources and social media streams has created new sources for geographers to increase their understanding of geographic phenomenon. However, the volume, velocity, and variety of these datasets combined with archived geographical data have pushed the limits of traditional systems in management and optimal performance. In order to address these needs, we have proposed PlanetSense, a novel distributed processing system that provides end-to-end functionality, from gathering raw data to generating actionable insights for geospatial intelligence in real-time. This architecture is scalable to support incremental data collection as well as integration with disparate and legacy data sources. It has built-in mechanism to perform machine learning and data mining algorithms on the processed data, allowing users to generate actionable intelligence in real-time. The presentation module displays easy to understand information through dynamic web interfaces and interactive querying systems. The case studies examining special event population during a collegiate football season and occupancy analysis of museum visitors have underpinned the need of deploying such a system for geospatial intelligence. Currently, we are working on developing software tools and expecting the platform to be fully operational in the next few months.

## 6. ACKNOWLEDGMENTS

This manuscript has been authored by employees of UT-Battelle, LLC, under contract DE-AC05-00OR22725 with the U.S. Department of Energy. Accordingly, the United States Government retains and the publisher, by accepting the article for publication, acknowledges that the United States Government retains a non-exclusive, paid-up, irrevocable, worldwide license to publish or reproduce the published form of this manuscript, or allow others to do so, for United States Government purposes.